# Optimizations of Autoencoders for Analysis and Classification of Microscopic In Situ Hybridization Images

*Authors: Aleksandar A. Yanev[1], Galina D. Momcheva[1,3], Stoyan P. Pavlov[2,3]*

[1]High School of Mathematics "D-r Petar Beron", Varna, Bulgaria
[2]Department of Anatomy and Cell biology, Medical University of Varna, Varna, Bulgaria
[3]Advanced Computational Bioimaging, Research Institute Medical University "Prof. Dr Paraskev Stoyanov" Varna, Bulgaria
E-mails: aleksandar.yanev.2004@gmail.com galina.momcheva@mgberon.com stoyan.pavlov@mu-varna.bg

***Abstract:*** *Currently, analysis of microscopic In Situ Hybridization images is done manually by experts. Precise evaluation and classification of such microscopic images can ease experts' work and reveal further insights about the data. In this work, we propose a deep-learning framework to detect and classify areas of microscopic images with similar levels of gene expression. The data we analyze requires an unsupervised learning model for which we employ a type of Artificial Neural Network — Deep Learning Autoencoders. The model's performance is optimized by balancing the latent layers' length and complexity and fine-tuning hyperparameters. The results are validated by adapting the mean-squared error (MSE) metric, and comparison to expert's evaluation.*

## 1. Introduction

In Situ Hybridization (ISH) is a method for the recognition and localization of specific nucleotide sequences in the nucleic acids (DNA and RNA) in cells and tissues [1]. Applications of this technique on microscopic images include the identification of infectious diseases (such as human immunodeficiency virus (HIV), herpes simplex virus (HSV), hepatitis B virus (HBV)), diagnosis and grading of cancer, cytogenetics, and analysis of gene expression. In the chromogenic variant of the reaction (chromogenic in situ hybridization – CISH), the hybridization is revealed by the production of a colored precipitate that can be easily observed, recognized, and documented using a standard bright-field microscopic imaging system [2]. The positive signals correspond to cells that actively produce (aka express) the product of a gene under investigation. [2,3]

In this study, we develop a workflow for automated, fast, reliable, and reproducible analysis of Chromogenic In-Situ Hybridization Images (CISH images). Currently, the substantial dependency of the method on tissue preparation, the conditions of hybridization and development of the color reaction, and the imaging parameters hinder the standardized analysis of large batches of such data. Thus, the "gold standard" for gene expression grading in CISH-stained tissue slides is expert assessment. This approach involves visual inspection of the slides or their images at various scales (Fig. 1) and manual labelling (as "positive" or "negative") or grading of the strength, density and/or distribution of the staining using more or less arbitrary ordinal scales for strength (e.g. "negative", "low", "moderate" and "strong"; or "-", "1+", "2+" and "3+") and patterns of expression ("ubiquitous", "regional" or "scattered") [2,4]

As mentioned above, we attempt to develop a reliable workflow using autoencoders. Autoencoders are a type of Artificial Neural Network used to learn efficient coding of unlabeled data. This unsupervised learning approach matches the profile of the examined data. The autoencoder consists of two main parts —the encoder maps the input into a code of representative features, while the decoder tries to recreate the original from this latent representation. As copying the input to the output is a common concern because it is indeed a valid solution (although not a useful one), autoencoders are usually forced to find and preserve the most representative features differently.

In this paper, we explore the properties of our workflow, such as methods used for extracting meaningful data for the algorithm's training, the length and complexity of the autoencoder used, its limitations and prospects for future improvements.

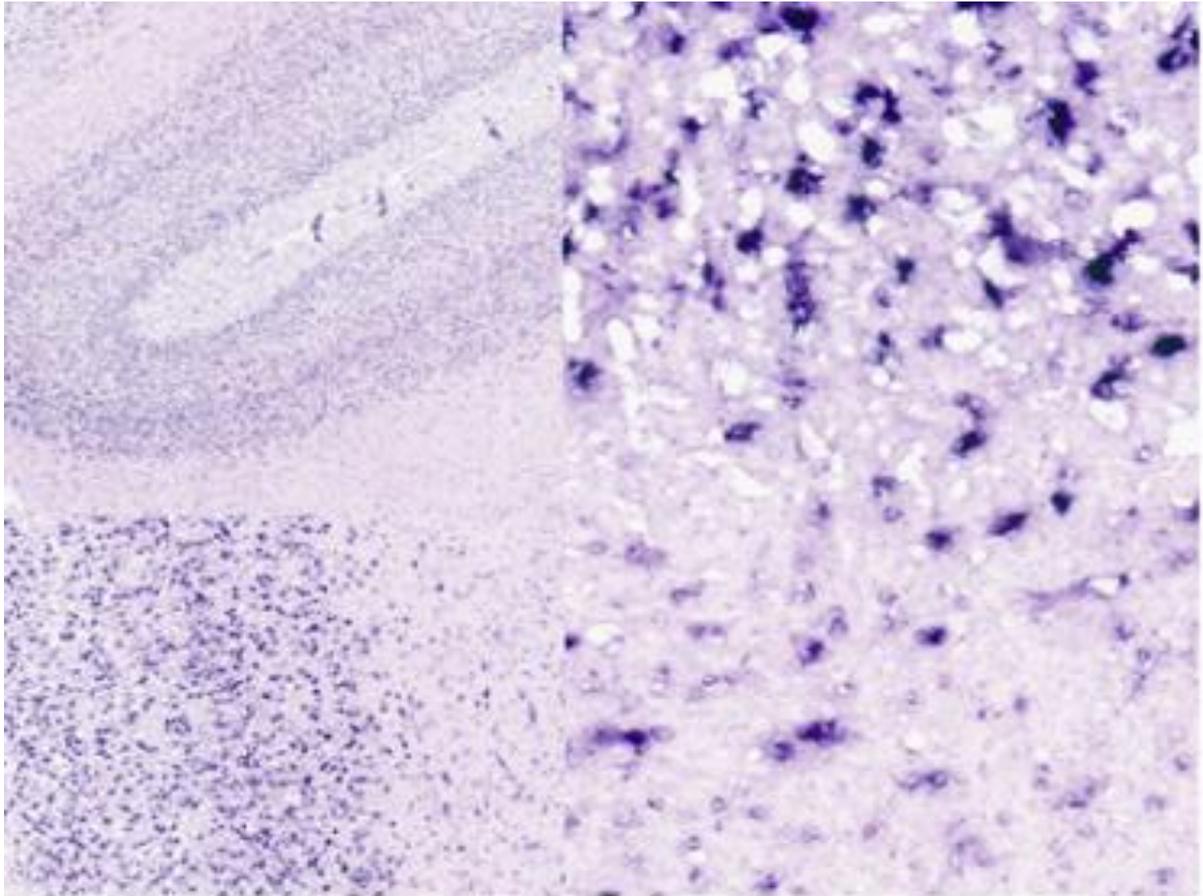
*Figure 1. CISH stained tissues are graded and evaluated at different scales.*

## 2. Extraction of meaningful data

Whole-slide CISH images' (Fig. 2A) sizes vary, but the microscopic image's dimensions usually span tens of thousands of pixels. As our goal is to classify areas of the image with similar gene expression levels, we must train the autoencoder on smaller regions (aka tiles) cut from the original. The choice of tile size depends on the goal and scale of the evaluation — e.g., entire slice, brain area, subregional evaluation, *et cetera*. In this study, we focused on the supracellular subregional level and thus have chosen square tiles with side 150 μm (at the scale of the original image — 0.5 μm/px, this corresponds to 300 px). Dividing the image into smaller tiles increases the possibility of introducing border artefacts due to random splitting of cells and areas with similar properties into different images (for example, cells may be cut in two, or the border of highly expressing region may be included in an image of a neighbouring area with lower levels of expression). Choosing an overlap between the tiles of 75 μm (150 px), we have ensured that each tile also will represent the transition between its adjacent tiles, and thus will reduce the possibility of border phenomena skewing the results.

    Whole-slide CISH microscopic images usually include a lot of background (areas without a tissue). These areas must be programmatically cut beforehand to reduce bias in the training/evaluation process. Furthermore, the precise removal of the background will reduce the computational time and the memory footprint of the autoencoder. In order to achieve this, we created masks of the whole-slide microscopic images beforehand that provide information about whether a certain pixel is part of the tissue or the background. (Fig. 2) For mask creation, we processed a lower-scale version of the whole slide image — gaussian blur, followed by an automatic triangle threshold and morphological fill holes. Usually, there are low-intensity imaging artifacts located outside of the tissue slice. To avoid the inclusion of the latter in the mask, we used a reconstruction from a seed created by a large-radius erosion of the mask image. In the final stage, we resize the thresholded image to the original scale, extract the coordinates where tissue is present, and define the tiles for these coordinates.

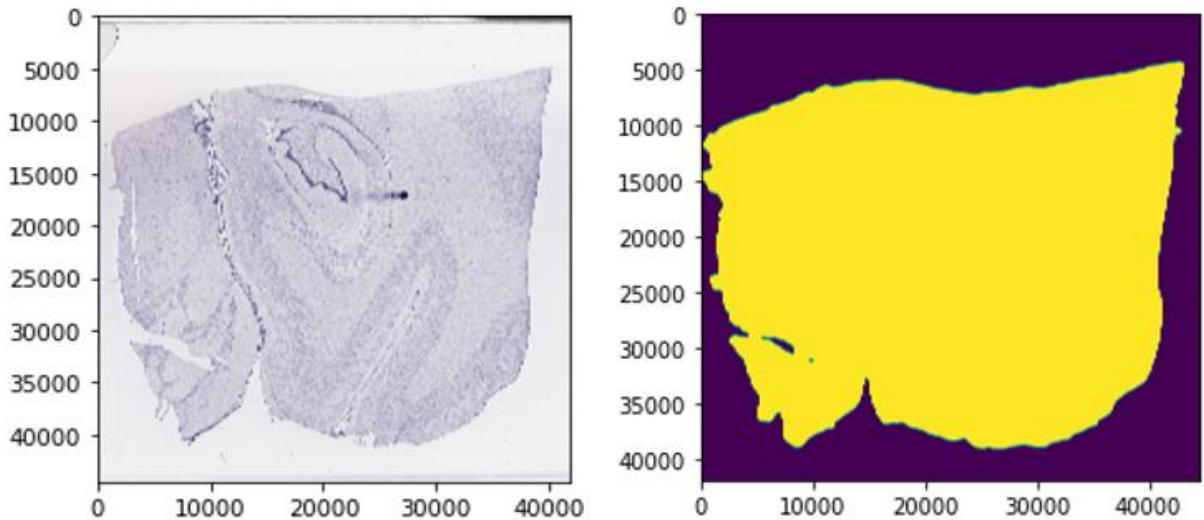

*Figure 2. Original CISH image (left) and its mask (right)*

## 3. The Autoencoder

3.1. Size of the encoder's output

One of the most important hyperparameters is the encoder's output size (the number of extracted features). Its dimensions are one of the deciding factors for the time/memory usage and the model's accuracy [5]. As we will see, we have chosen to reduce the 300x300px tiles to just 2 floating point numbers, which comes with its benefits and limitations.

In our workflow, we have prioritized reducing computational time without introducing major memory usage disadvantages. The current best model uses less than 2 hours to complete its training, classify all data, and reconstruct the initial images in a format that experts can evaluate. Another significant advantage is that with just two floating numbers, we can visualize thousands of tiles on 2D planes and search for meaningful correlations in the data. In this way, biology experts without a computer science background can easily look for new relations in available data.

However, we must note some of the limitations of the model. The major one is the inability of the model to actually "recreate" a given photo (Fig. 3). As the data we use for reconstruction (the latent layer of the autoencoder) is extremely limited, the model seems not to remember any spatial information, and the general space orientation is distorted. Thus, the decoder returns images that, despite representing similar intensity features, do not resemble the original picture and can not be easily understood by a human observer.

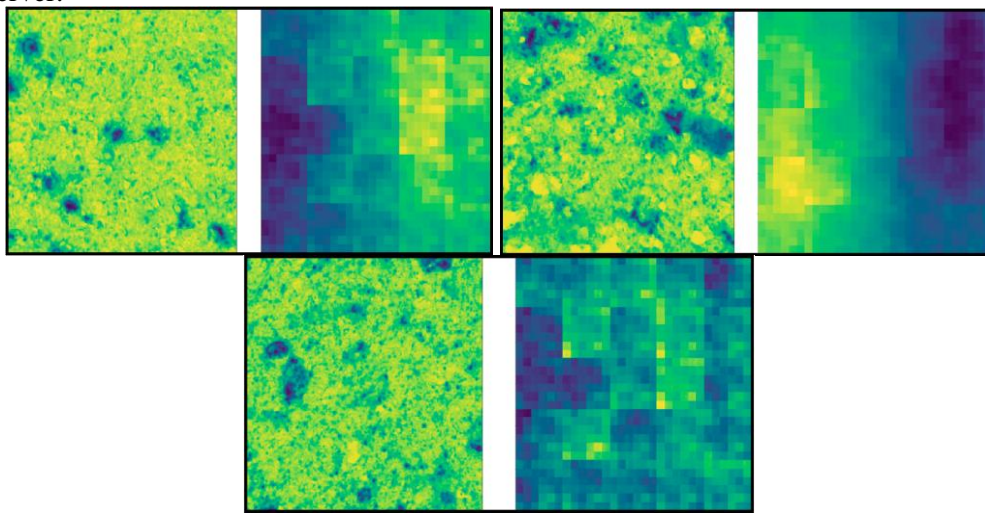

*Figure 3. Three original tiles and their respective reconstructions from the autoencoder*

3.2. Evaluation of loss and choice of an optimizer

We are performing unsupervised classification with the freedom to vary the number of assigned classes. Testing with two of the most common functions for image analysis, such as MSE (mean-squared error) and MAE (mean absolute error), shows that the MSE performs better (Fig 4). We must acknowledge that the current mask generation is not perfect, and some background tiles are indeed fed into the autoencoder. As significant outliers, they increase the value of the loss function — as we see in Fig. 4, the neural network converges around the value of 3. To demonstrate that this loss does not impede the correct classification of the tiles, we have performed two tests — a real-world and a programmatic one.

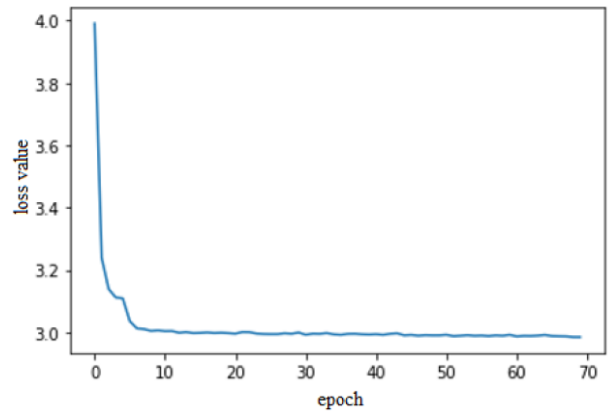

*Figure 4. Plot of the loss function with regard to epoch number*

The real-world test includes expert evaluation of the resulting whole CISH images and approval that the labels match the manual evaluation — areas of matching levels of gene expression belong in the same clusters appointed by the neural network.

The programmatic test used the preceding work of Pavlov S., Atanasov S., Momcheva G. [6,7], who extracted the exact coordinates for the two CISH images on which the autoencoder is trained. Evaluation of the autoencoder using the perfect dataset shows a convergence of the loss function at a similar value of 3, thus proving that the outlier background tiles do not change the result drastically. Furthermore, a comparison of the results reveals that after being trained on each data set, the resulting classification outlines the same regions as having similar levels of gene expression. These tests demonstrate the masking algorithm's ability to remove enough background data as well as the ability of our autoencoder to withstand small amounts of background data without bias in the results.

We must acknowledge that such ability is necessary because of the beforementioned differences between CISH images due to uncontrollable outside factors such as dependency on tissue preparation and conditions of the microscopic image acquisition.

The measurement of loss and the actual accuracy of the model is also closely related to the optimizer function. Qualities of our data, such as the evidence of rare features, strongly point to the usage of adaptive optimizers. In choosing an optimizer function, we have considered benchmarks of optimizers [8] concerning our data. After testing, although with a difference of less than 0.02, the Adam optimizer [9] outperforms RMSprop [10], Adagrad [11], and Adadelta [12].

3.3. Batch sizes and epochs

Batch sizes are usually a predefined parameter that sets the number of data points — in our case, tiles — needed before an update of the neural network's parameters. Larger batch sizes reduce computational time as parameters update less frequently, but when combined with the number of epochs, disbalance may lead to underfitting of the model. We have found no significant difference in convergence when batch sizes are between 30 and 60 combined with 70 epochs for the autoencoder training.

3.4. Layers

One of the main problems encountered was the incapability of the model to extract meaningful features. The solution was the continued deepening of the model—14 blocks of layers in total (Fig. 5). By looking at results from an autoencoder with similar architecture but only 8 layers (4 for encoder and 4 for decoder - 2 convolutional and 2 linear layers each), we observed that the network still manages to extract features but not as complex or as well represented as needed. In conclusion, the model we have decided on has an encoder that chronologically uses 4 blocks of Convolution and MaxPooling with ReLU activation function, followed by 3 blocks of Linear with LeakyReLU activation function. The decoder uses 3 blocks of Linear with LeakyReLU activation function, followed by 3 blocks of Transpose Convolution with ReLU activation function and one with Sigmoid activation function. As the 'flatten' and the 'unflatten' functions just ease our work by changing the data dimensions, we represent them on the figure but do not consider them in counting neural network depth. In the following paragraphs, we will discuss the purpose and properties of these blocks.

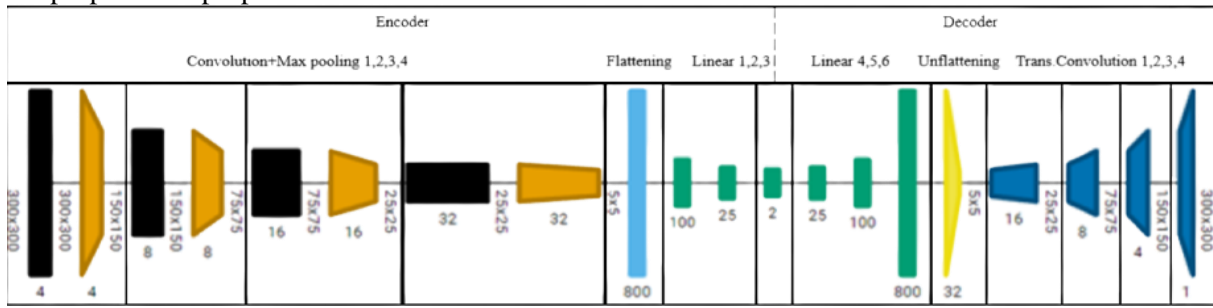

*Figure 5. Scheme of the final autoencoder created with Graphiz [13]—Encoder with four layers of Convolution combined with Max Pooling, Flatten layer, three Linear layers and decoder with three linear layers, Unflatten layer and four Transpose Convolution layers*

3.4.1. Linear blocks

Linear blocks are the simplest hidden layer. The purpose of this block is mainly to reshape the data while still preserving features by performing the linear transformation from (1). The activation functions [14] used afterward enable the linear blocks to learn. Activation functions are predefined mathematical expressions that consider the result of a specific neuron in the network and its bias and decide whether or not to fire the said neuron.

(1) $$y = xA^T + b$$

The problem we encountered was the effectiveness of the activation functions. Initially, a compromise between ReLU and Sigmoid functions was used and provided semi-correct results, but after some testing, we encountered the so-called "Dying ReLU" problem [15] that is connected to a number of neurons never firing and thus acting as dead weight. The replacement of normal ReLU-s with Leaky ReLU-s seems to resolve the issue.

In the encoder, we use three linear layers that gradually reduce the values remaining from the one tile from 800 to just 2 — the gradual steps are, in order, from 800 to 100, from 100 to 25, and from 25 to 2. Distributing the size reduction into three layers allows for better features as each neuron's significance is calculated on multiple levels and therefore is more evident.

The decoder uses the same three linear layer blocks with reversed dimensions — from 2 to 25, from 25 to 100, and from 100 to 800. In conclusion of the linear layers, we must mention the usage of "flatten"- and "unflatten"-functions. The flatten function transforms the data from the initial two-dimensional tile to a one-dimensional array. Similarly, the unflatten layer uses the one-dimensional output provided after the linear transformations of the decoder to present the data as a tile again.

3.4.2. Convolution and max pooling

Convolution [16] is a popular technique used for image recognition and processing that is specifically designed to process pixel data. Convolutional layers are quite complex because the output shape is affected by the shape of its input and the choice of parameters such as kernel shape, zero padding, and strides. The relationship between these properties is not trivial to infer. This difficulty contrasts with fully-connected or linear layers, whose output size is independent of the input size. Additionally, convolutional layers also feature a pooling stage, adding yet another level of complexity with respect to fully connected/linear layers.

As convolution creates fields with very similar pixel values, we use a method called pooling to extract sharper features and scale down the image to reach a usable encoder output. As we want to extract the most significant levels of gene expression in a particular tile, max-pooling is suitable to pass down the information.

The pixels most significant for estimating the tile's features may be disregarded if the convolution kernels are too big—a single high-value cell may influence the categorization of an entire region. That is why in order to have balance and retain the true values even after convolution, we have decided to use four layers with smaller kernel sizes, each followed by a ReLU function (in this case we have not seen problems with the standard ReLU activation function as above) and a max pooling layer. After continuous testing and tweaking of the parameters of the model, the three convolutions have blocks as follows:

Table 1. Convolution block layers

| Block number | Convolution kernel size (in_channels, out_channels, kernel_size, padding=1) | Activation function | Max pooling layer size |
|---|---|---|---|
| 1 | (1,4,3) | ReLU | (2,2) |
| 2 | (4,8,3) | ReLU | (2,2) |
| 3 | (8,16,3) | ReLU | (3,3) |
| 4 | (16,32,3) | ReLU | (5,5) |

The decoder's activation functions and the Transpose Convolutional layers ( the reverse function of the Convolution layers) use the same parameters as the encoder track. We must mention two technical details: the absence of any form of pooling in the decoder and the presence of a sigmoid function instead of a ReLU function in the last Transpose Convolution. This modification prevents significant jumps in output values that may disrupt the clustering we perform later.

Without convolution, the neural network is unable to learn any significant features or even converge, as it can be seen in the loss function in Fig. 6

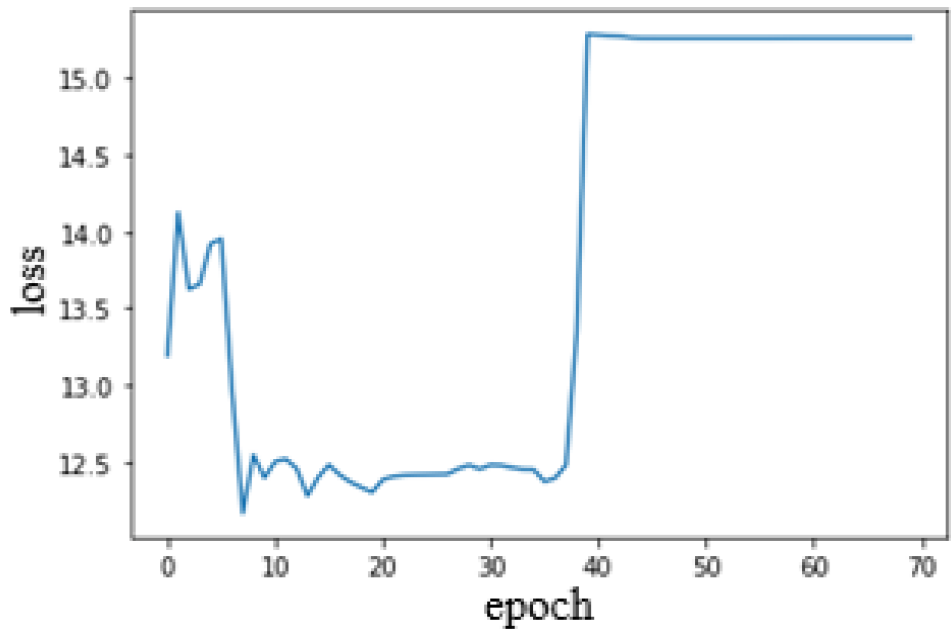

*Figure 6. Loss function of autoencoder model without convolution*

4. Clustering of results

The autoencoder provides two floating point values that represent the initial tile. In order to cluster the tiles and examine any relationships between them, we must use a clustering algorithm. Hard and fuzzy clustering algorithms are the two most common types of such algorithms. As the gene expression in our tiles is not defined into clearly separate classes — like a dog or cat guesser — we have decided to use Fuzzy c-means clustering. [17, 18] In short, the step-wise approach of the Fuzzy c-means clustering goes as follows:

- fix the number of clusters, let it be c, and select a fuzziness parameter, let it be m (generally 1.25<m<2), and initialize a partition matrix U $[u_{i,j}]$
- Then repeatedly do the following:
1. Calculate cluster centers from (2) (centroids)

(2)
$$c_{i,j} = \frac{\sum_{i=1}^{n} u_{i,j}^m * x_i}{\sum_{i=1}^{n} u_{i,j}^m}$$

2. Update U from (3) (centroids)

(3)
$$u_{i,j} = \frac{1}{\sum_{k=1}^{c} \frac{|x_i-c_j|^{\frac{2}{m-1}}}{|x_i-c_k|^{\frac{2}{m-1}}}}$$

3. stop process when U stops changing significantly or when desired

Fig. 7 shows the number of centers and the value of the associated FPC (fuzzy partition coefficient; a metric for the model performance describing the data). In our results, FPC decreases with the number of clusters. This behavior is expected as there are no apparent clusters in our data — the tiles form a cloud that demonstrates an almost continuous gradient in the calculated features.

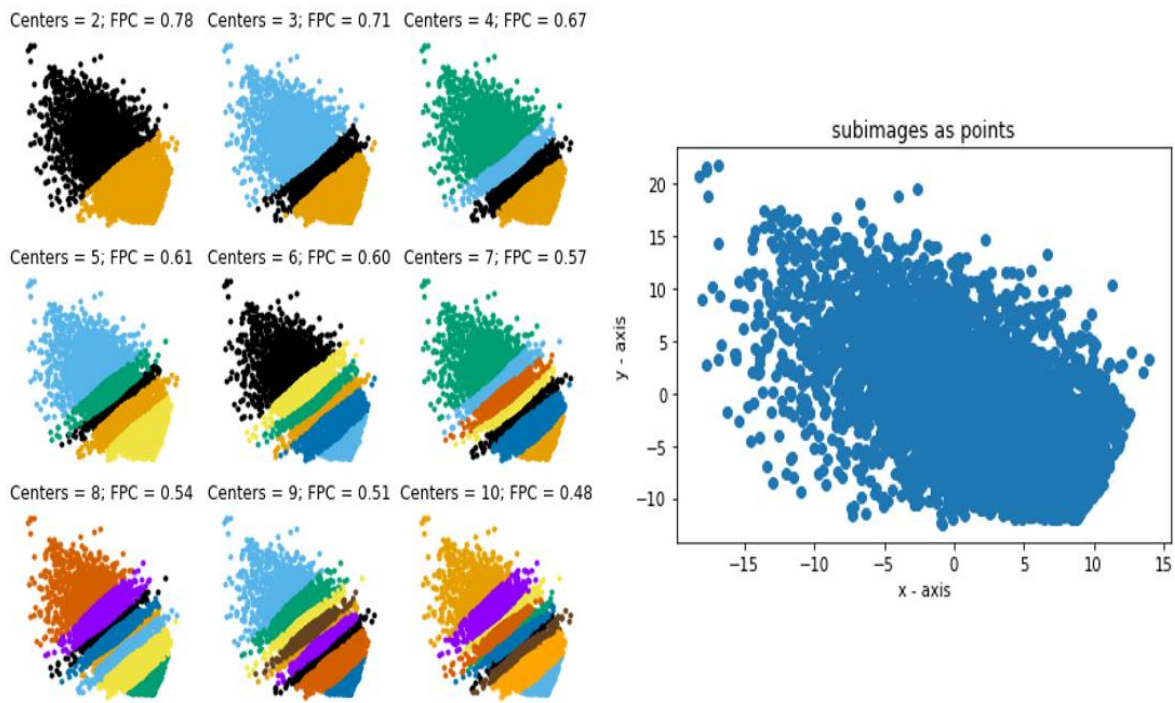

*Figure 7. The distribution of tiles on the 2D plane and their clustering depending on the number of centroids*

## 5. Analysis of the results

In order to manually inspect the results, we examine the 7 clusters model (Fig. 8) and reconstruct the starting images by returning each tile (colored coded by class) to its place in the original microscopic image. (Fig. 9) We have chosen seven clusters to recreate a real-life evaluation—as the number of clusters increases, it becomes much harder to classify the tiles correctly by hand.

The apparent linearity in the two-dimensional feature space (Fig. 8) transfers into the well-visible correspondence between the colored-coded classes and regions with particular gene expression patterns in the unprocessed images (Fig. 9). The reconstructions outline and demonstrate that the algorithm classifies regions with similar staining properties in the two images to the same class.

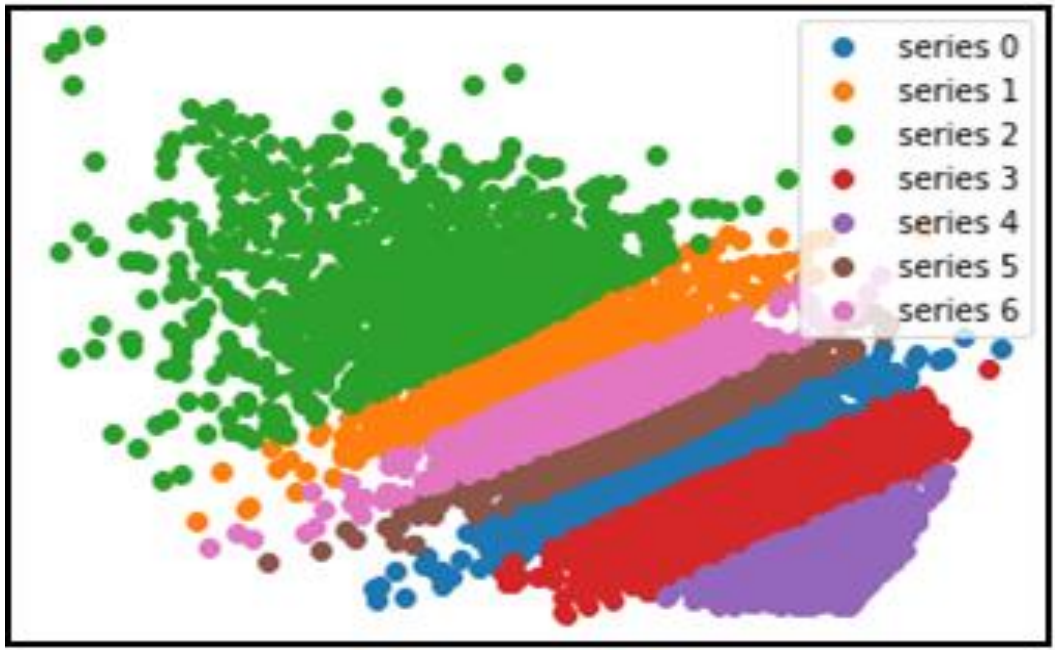

*Figure 8. Clustering of tiles with seven centroids*

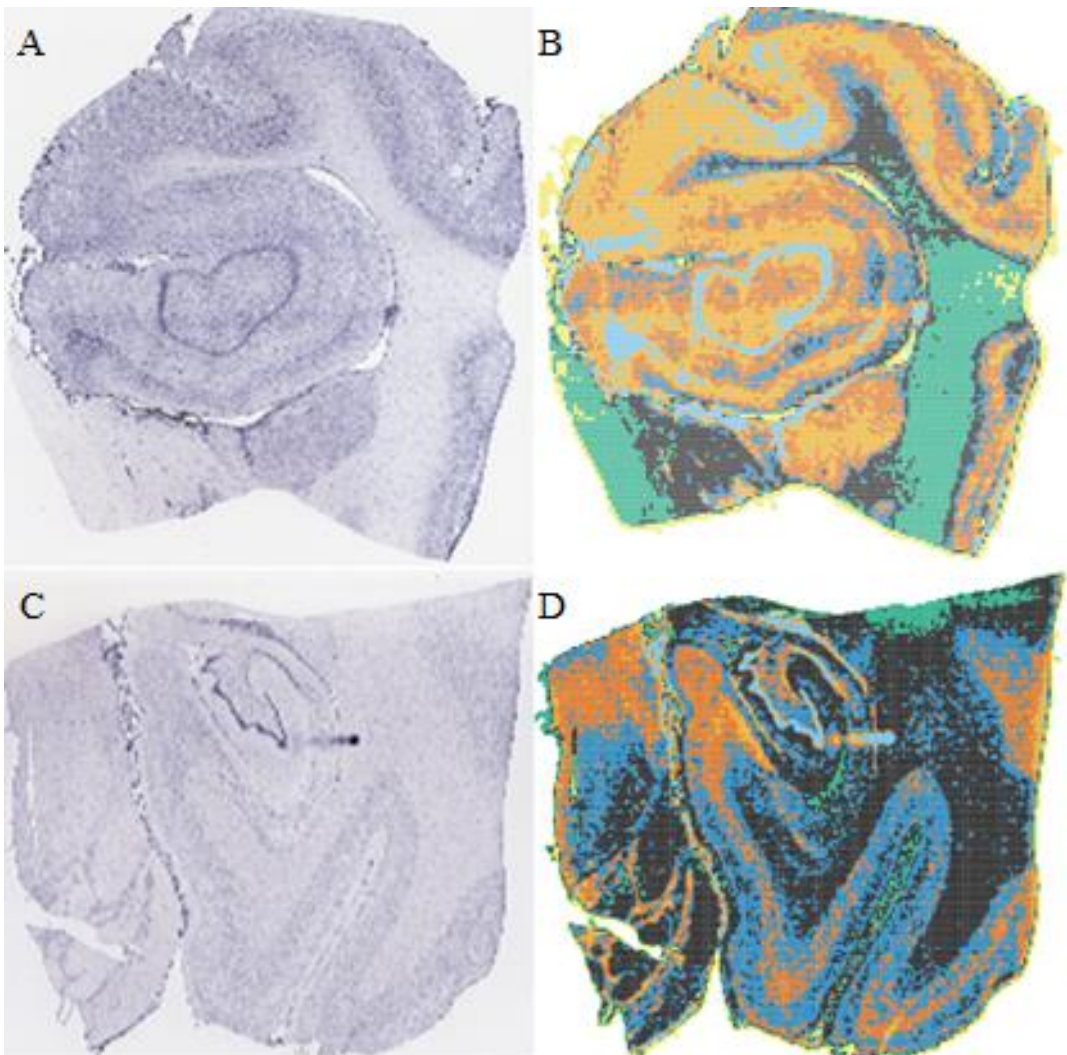

*Figure 9. Original images (A,C) and their color-coded reconstruction (B,D) with classified tiles*

## 6. Future plans and prospects

In this paper, we present a concept that an autoencoder latent layer can be used as a feature space for unsupervised classification of the staining patterns in microscopic CISH images. For more consistent results in the future, the autoencoder behavior must be studied while training with more diverse and representative batches of data. To generalize the algorithm and to account for differences in the conditions of hybridization and image exposure, the autoencoder must be trained on a larger number of random tiles from more images.

Furthermore, the masking algorithm may be improved by implementing a more complex algorithm to reduce the tiles that do not contain part of the microscopic image, thus reducing computational time.

Also, we plan to develop an interface for the definition of either random selection of training tiles or predefined coordinates of a region of interest for analysis; an interface for the definition of scale i.e. size of the processed tiles; strategies for the deployment of the training algorithm and the trained network and design of a user-friendly interface for practical application. We plan to gradually increase the number of features and look for any improvements and changes in the usefulness of the described approach. For example, the decoder portion of a similar autoencoder with feature-rich latent layer may be combined with a model for the random generation of feature values and included in a GAN for data augmentation.

As the evaluation method of other microscopic images, when boiled down to its essence, is very similar, it will be fascinating to see how the model will behave when applied to different microscopic and macroscopic images.